\documentclass[preprint,12pt]{elsarticle}
\usepackage[margin=2.5cm]{geometry}
\usepackage{amssymb}
\usepackage{siunitx}
\usepackage{lineno}
\usepackage{blindtext}
\usepackage{setspace}
\usepackage{hyperref}

\date{}
\begin{document}

\begin{frontmatter}
{
\title{Deformation-induced homogenization of the multi-phase senary high-entropy alloy MoNbTaTiVZr processed by high-pressure torsion}

\author[affiliation_a]{Chuyi Duan\corref{cor1}}
\cortext[cor1]{Corresponding author.}
\ead{chuyi.duan@tum.de}

\author[affiliation_b]{Aleksander Kostka}
\author[affiliation_c]{Xiaohu Li}
\author[affiliation_d]{Zirong Peng}
\author[affiliation_e]{Peter Kutlesa}
\author[affiliation_e]{Reinhard Pippan}
\author[affiliation_a]{Ewald Werner}

\affiliation[affiliation_a]{organization={Institute of Materials Science, Technical University of Munich},%Department and Organization
            addressline={Boltzmannstr. 15}, 
            city={Garching},
            postcode={85748}, 
            %state={},
            country={Germany}}
\affiliation[affiliation_b]{organization={Center for Interface-Dominated High Performance Materials (ZGH)},%Department and Organization
            addressline={Universitätsstr. 150}, 
            city={Bochum},
            postcode={44780}, 
            %state={},
            country={Germany}}
\affiliation[affiliation_c]{organization={Heinz Maier-Leibnitz Zentrum (MLZ), Technical University of Munich},%Department and Organization
	addressline={Lichtenbergstr. 1}, 
	city={Garching},
	postcode={85748}, 
	%state={},
	country={Germany}}
\affiliation[affiliation_d]{organization={Carl Zeiss Microscopy GmbH},%Department and Organization
            addressline={Kistlerhofstr. 75}, 
            city={München},
            postcode={81379}, 
            %state={},
            country={Germany}}       
\affiliation[affiliation_e]{organization={Erich Schmid Institute of Materials Science, Austrian Academy of Sciences},%Department and Organization
            addressline={Jahnstr. 12}, 
            city={Leoben},
            postcode={8700}, 
            %state={},
            country={Austria}}

\begin{abstract}
Dendritic microstructures are frequently observed in as-solidified refractory high-entropy alloys (RHEAs), and their homogenization typically requires a long-term heat treatment at extremely high temperatures. High-pressure torsion (HPT) has been shown to be capable of mixing immiscible systems at room temperature, and therefore represents a promising technique for homogenizing dendritic RHEAs. In this work, the as-solidified RHEA MoNbTaTiVZr was processed up to 40 revolutions by HPT. It was found that the dendritic microstructure was eliminated, resulting in a chemical homogeneity at a von Mises equivalent shear strain of about 400. The study of deformation mechanism showed an initial strain localization, followed by a co-deformation of the dendritic and interdendritic regions. In the co-deformation step, the Zr-rich interdendritic region gradually disappeared. The deformation-induced mixing also led to the formation of an ultra-fine grained (UFG) microstructure, exhibiting a grain size of approximately 50 nm. The microhardness increased from 500 HV in the as-solidified to 675 HV in the homogenized UFG state. The underlying mechanisms responsible for the microhardness enhancement, such as grain refinement and solid solution strengthening, were also discussed.
\end{abstract}

\begin{keyword}
High-entropy alloy \sep Severe plastic deformation \sep Chemical homogenization
\end{keyword}

}

\end{frontmatter}
\vspace{-21.5cm}
\textbf{\underline{Published in}\\
Mater. Sci. Eng. A 871 (2023) 144923\\ 
DOI: \url{https://doi.org/10.1016/j.msea.2023.144923}}
\newpage
%% main text
\section{Introduction}
\label{intro}
High entropy alloys (HEAs) which consist of multiple principal elements have attracted tremendous research attention due to their superior mechanical properties \cite{Zhang2014, Miracle2017}. 
The high configurational entropy in HEAs makes it thermodynamically preferable to form single-phase alloys \cite{Yeh2004}. 
However, with the addition of alloying elements, not only the configurational entropy is increased, but also the interactions between elements in the alloy become more complex, which may lead to the formation of additional phases \cite{Senkov2015}. 
The rapidly developing aerospace industry requires increasingly high performance of materials at high temperatures, and these requirements are already difficult to meet with conventional Ni-based superalloys \cite{Senkov2018}. Refractory HEAs (RHEAs) are gradually coming into attention as high temperature materials, with their main elements being predominantly refractory elements. 
The complexity of the elemental composition in RHEAs frequently leads to chemical segregation during solidification from the liquid-state \cite{Senkov2018}. This is often due to the large differences in the melting points of the different elements and their respective effects on liquid phase stabilization \cite{Gao2016}. As a result, as-solidified RHEAs are often dendritic. There are different opinions on the phase composition of RHEAs containing dendrites. On the one hand, because only one phase peak is often seen in XRD, dendritic RHEAs are often considered to be single-phase as well, such as the original RHEA MoNbTaW proposed by Senkov \cite{Senkov2010}. On the other hand, Zhang et al. suggest that this chemical segregation is actually due to the existence of two phases \cite{Zhang2017}. 
\par
High-pressure torsion (HPT) is a well-established technique of severe plastic deformation (SPD). It has been widely employed to study the strain hardening and grain refinement behavior of metallic materials, including single-phase HEAs \cite{Schuh2015,Schuh2018,Tang2015}. Furthermore, dual-phase HEAs have also been investigated extensively due to their unique mechanical and physical properties. For example, some dual-phase HEAs consist of both body-centered cubic (BCC) and face-centered cubic (FCC) phases \cite{Taheriniya2021}, while others comprise a soft solid solution matrix and hard intermetallic phases \cite{Edalati2021Carbide}. In addition, the use of HPT has been extended for the fabrication of HEAs by mechanical alloying pure elemental powders \cite{Kilmametov2019}. Moreover, HPT has recently been used to study the forced formation of supersaturated solid solutions in immiscible systems, which has attracted considerable attention due to the potential for discovering new materials with improved properties \cite{Bachmaier2014, Kormout2017}. The possibility of the forced formation of solid solution in immiscible systems using HPT brings up an important inquiry regarding the effectiveness of HPT-processing in eliminating the elemental segregation and achieving chemical homogeneity in RHEAs. Hence, a thorough investigation is necessary to shed light on this topic and advance our understanding of the properties and behavior of these materials.
\par
This work focuses on the elemental mixing mechanism during high-pressure torsion of a multi-phase, dendritic as-solidified, senary refractory high entropy alloy of MoNbTaTiVZr. The evolution of the microstructure, hardness and composition during the HPT enables to identify mechanisms governing the elemental mixing. 
\section{Experimental}
\label{exper}
% Sample preparation arc melting
The equimolar RHEA MoNbTaTiVZr was prepared by arc melting of pure elements (purity higher than 99.5 wt.\%) on a water-cooled copper hearth in an arc-melter (MAM-1, Edmund Bühler). The ingot was flipped and remelted at least 10 times to ensure chemical homogeneity. 
Before melting, the chamber of the arc-melter was evacuated and backfilled with high purity Argon (5N) for 5 times. A Ti-getter was used to further remove the oxygen. Disk-shaped samples with a diameter of 6 mm and a thickness of 0.8 mm were sliced from the ingot using electric discharge machining and a wire diamond saw. More details of the sample preparation can be found in \cite{Duan2022}. 
\par
\begin{figure}[!ht]
	\centering
	\includegraphics[width=0.5\textwidth]{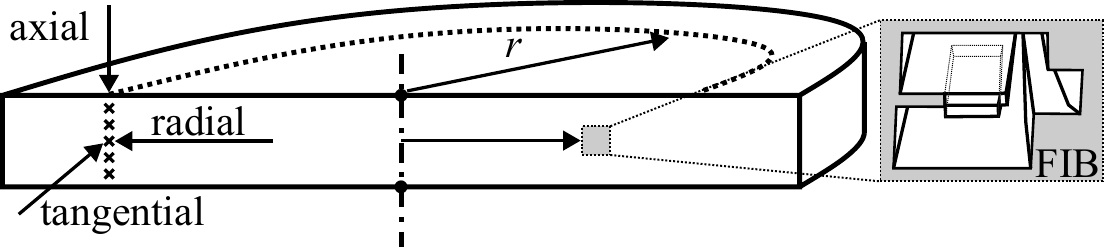}
	\caption{A sketch showing the position of microhardness measurements and the TEM specimen prepared by FIB. }
	\label{fig:HPT_direction}
\end{figure}

% High-pressure torsion
The disks were subjected to HPT-processing using conventional Bridgman-anvils with a cavity height of 0.6 mm, under a quasi-hydrostatic stress of 7 GPa, for varied numbers of revolution $N$: 0.25, 2, 4, 10, 20, 30 and 40. The samples after HPT-processing will be denoted as $N$0.25, $N$2, $N$4, $N$10, $N$20, $N$30 and $N$40, respectively. The rate of processing was chosen as 0.2 revolutions per minute to avoid excessive generation of heat. A disk which was compressed without introducing shear strains, denoted as $N$0, was used for comparison. The HPT-disks were divided in half, and the surface defined by the axial and radial directions was examined using microscopic characterization, as shown in Fig. \ref{fig:HPT_direction}. 
\par
% Microstructure SEM, FIB, TEM 
Scanning electron microscopy (SEM) was conducted in the microscope JEOL JSM-6490 equipped with an energy-dispersive X-ray spectrometer (EDS) operated at 20 kV. The HPT-disks as well as as-solidified specimens were embedded, mechanically ground and finally polished with OPS (oxide polishing suspension, 0.25 \textmu m). 
\par
Transmission electron microscopy (TEM) was carried out in an aberration corrected microscope JEOL JEM-ARM200F operated at 200 kV. 
The TEM specimens were prepared using the focused ion beam (FIB) method in an FEI Helios G4 CX operated at 30 kV. 
Lamella specimens with a dimension of about 8$\times$8$\times$1 \textmu m were extracted from the radial direction of HPT-disks using $\mathrm{Ga}^{+}$ ion milling. A carbon protective deposition of about 3 \textmu m was sputtered prior to thinning process. During the final step of sample thinning, the accelerating voltage was reduced to 8 kV in order to minimize the beam damage. 
\par
% SXRD and Lab XRD
Synchrotron high-energy X-ray diffraction (HEXRD) as well as laboratory XRD analysis were used to examine the phase composition. HEXRD was performed in transmission mode with a wavelength $\lambda$ of 0.14235 \AA\,(photon energy about 87 keV) and a wavelength of 0.124 \AA\,(photon energy about 100 keV) at the High Energy Materials Science (HEMS) beamline at PETRA III at DESY in Hamburg, Germany. The beam size was set as 0.7$\times$0.7 mm by adjusting the slit aperture. A PerkinElmer XRD 1622 flat panel detector was placed at a distance to sample of about 1400 mm to record the diffracted beam. Standard $\mathrm{LaB_{6}}$ powder was used to calibrate the results. The recorded 2D diffraction patterns were integrated in the software FIT2D \cite{Hammersley1996}. Rietveld refinement was conducted in the software MAUD \cite{Lutterotti1999}. Laboratory XRD was performed in a Brucker D8 diffractometer using Cu as anode material (wavelength $\lambda$ 1.5406 \AA) with a beam size of 0.5$\times$0.5 mm. Due to the different wavelength of XRD applied in this paper, diffraction intensity was plotted against the wave vector $Q=4\pi \mathrm{sin}\theta/\lambda$ instead of the diffraction angle $\theta$ for easier comparison. 
\par

Vickers microhardness measurements were carried out using a Reichert-Jung micro-DUROMAT 4000E with a load of 100 g and a dwell time of 10 s. The results were averaged from five measurements along the axial (thickness) direction taken from the polished surface defined by the axial and radial directions. 
\section{Results}
\label{sec:results}
\subsection{Initial microstructure prior to HPT-processing}
\begin{figure}[!ht]
	\centering
	\includegraphics[width=\textwidth]{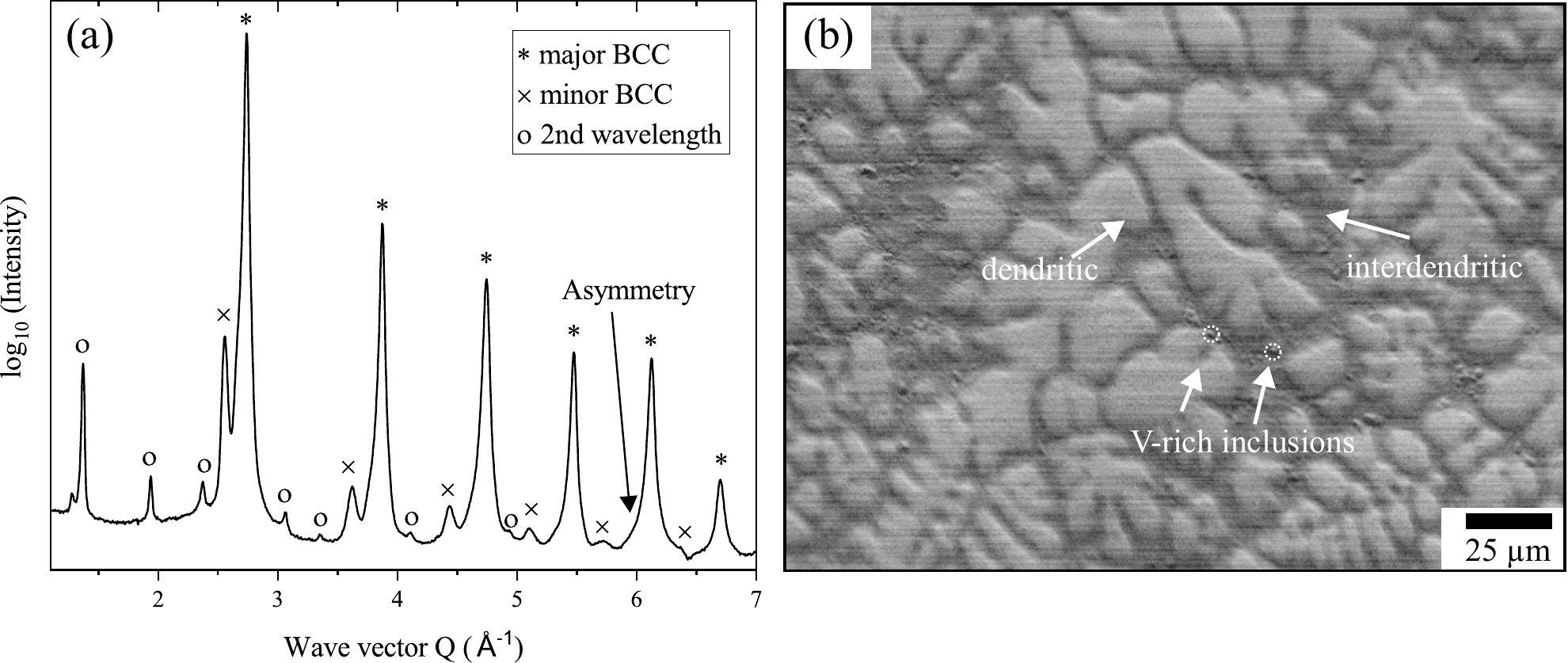}
	\caption{(a) The X-ray diffractogram at $\varepsilon\sim$ 0 in an wave vector-log$_{10}$(intensity)-plot. (b) SEM micrograph of the as-solidified RHEA.}
	\label{fig:Before_HPT}
\end{figure}

Fig. \ref{fig:Before_HPT} shows the XRD pattern and SEM micrograph of the dendritic microstructure of the as-solidified equimolar RHEA MoNbTaTiVZr. Two groups of BCC peaks with a pronounced asymmetry to the peaks of the major BCC phase are also visible. According to our previous study \cite{Duan2022}, the RHEA has a grain size of approximately 100 \textmu m and consists of dendritic regions depleted of Zr and interdendritic regions enriched in Zr. 
\par
The SEM micrograph at large magnification (\ref{fig:V_influence}a in the appendix) shows the existence of V-rich inclusions with a size of about 500 nm to 1 \textmu m distributed in the interdendritic region. An additional investigation on the influence of the alloying element V was conducted. 
By analyzing their diffraction peaks obtained from HEXRD (Fig. \ref{fig:V_influence}b in the appendix), the peaks of the minor phases left to the major BCC phase in the diffractogram are proved to belong to V-rich inclusions. 
Due to the low peak intensity of the minor phase, the diffraction intensity of RHEA was plotted logarithmically (Fig \ref{fig:Before_HPT}a) to allow an examination of the minor phase. 
Note that the synchrotron X-ray beam contains a second wavelength of about 10\% total intensity. The resulting peaks are marked with "$\mathrm{o}$" in the diffractogram. 
The plot shows that the V-rich inclusions also have a BCC crystal structure. In addition, an asymmetry on the left side of the major BCC phase peak was observed. The asymmetry is more obvious at large wave vectors Q (indicated by an arrow), which may suggest the existence of an additional group of peaks. By Rietveld refinement (Fig. \ref{fig:AsymRR}, appendix), the additional group of peaks could significantly reduce the difference between the calculated profile and the experimental result. 
The additional peaks are assigned to the interdendrtic region and are also indexed as BCC phase as the asymmetry lies always between the peaks of the major and the minor BCC phases. It is assumed that the peaks of the dendritic and interdendritic regions are so close to each other that they cannot be distinguished in the diffractogram. Despite the influence of the V-rich inclusions distributed in the interdendritic region, the SEM-EDS result in Tab. \ref{tab:SEM-EDS} can still provide a rough estimate of the chemical composition. Note that results of the inderdendritic region are obtained from positions containing as few V-rich inclusions as possible. 
\par
\begin{table}[!ht]
	\centering
	\caption{Concentration of chemical elements obtained from SEM-EDS measurements (in atomic percent) of the dendritic (D) and interdendritic (ID) regions.}
		\begin{tabular}{c c c c c c c}\hline
			Region & Mo & Nb & Ta & Ti & V & Zr \\ \hline
			D & {21.1$\pm$0.5} & {20.3$\pm$0.1} & {21.3$\pm$0.2} & {14.4$\pm$0.2} & {14.3$\pm$0.2} & {8.8$\pm$0.2}	 \\ 
			ID & {7.5$\pm$2.2} & {12.2$\pm$1.4} & {4.9$\pm$1.5} & {17.6$\pm$0.6} & {13.9$\pm$1.8} & {43.8$\pm$7.1}	 \\ \hline
	\end{tabular}
	\label{tab:SEM-EDS}
\end{table}
\subsection{Characterization of the HPT-processed RHEA}
\subsubsection{Microhardness evolution}
The von Mises equivalent shear strain at a position $r$ of HPT-disks can be stated as 
\begin{equation}
	\varepsilon = \frac{2\pi N r}{\sqrt{3}t}, \; 0\leq r \leq R,
	\label{eq:epsilon}
\end{equation}
where $N$ is the number of revolutions, $R$ is the radius of the disk and $t$ is its thickness after HPT \cite{Valiev2000}. Due to the uncertainties of the sample dimension, e.g. inexact $r$ after grinding and polishing in metallographic preparation and small variations of $t$ along the radius, the strain value will be shown as an approximate value along with "$\sim$" in this study. 
\par
\begin{figure}[!ht]
	\centering
	\includegraphics[width=0.75\textwidth]{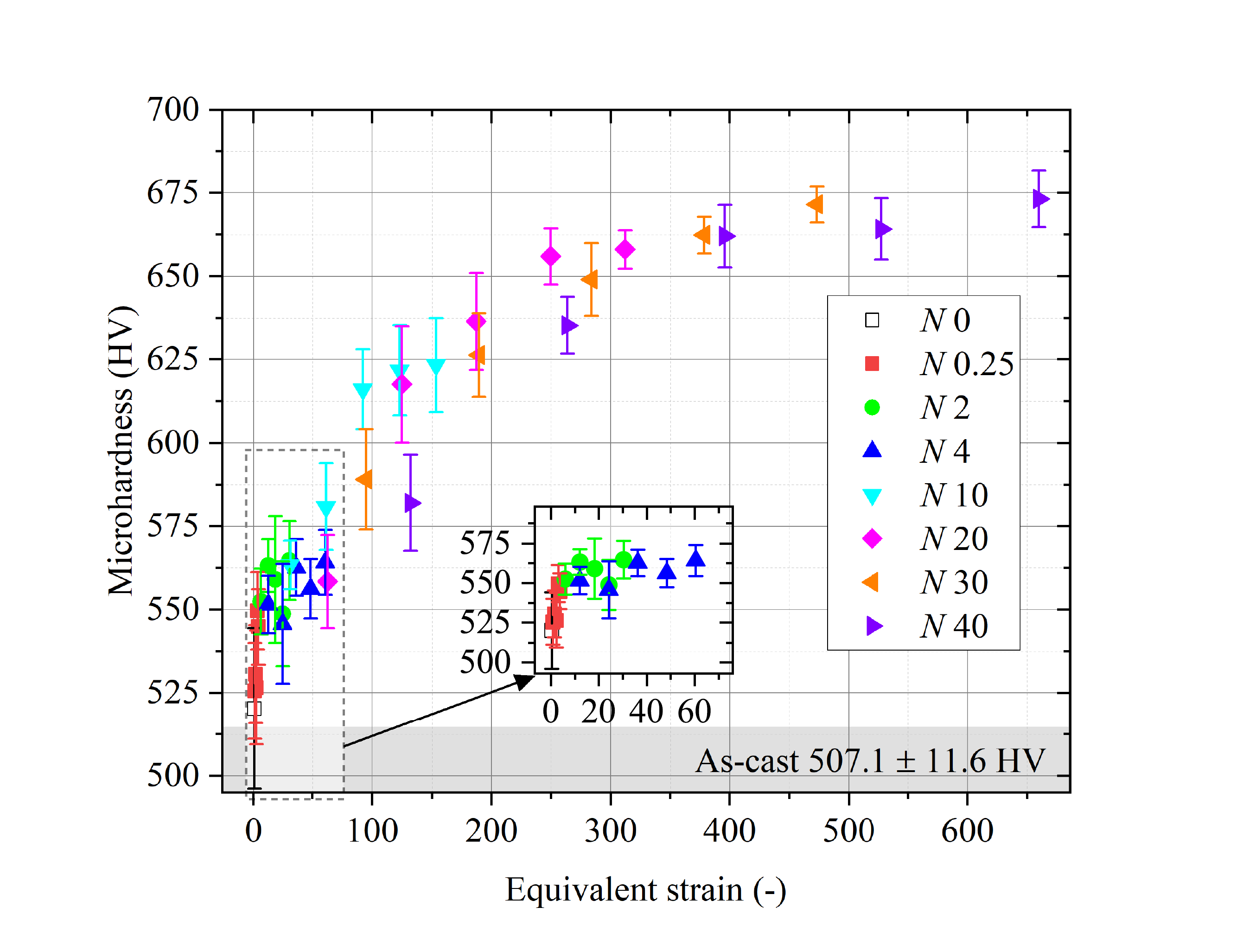}
	\caption{Microhardness evolution during HPT-processing. The inset shows a detailed view up to the strain of about 70. }
	\label{fig:microhardness}
\end{figure}

Fig. \ref{fig:microhardness} shows the microhardness evolution with increasing applied strain. Note that the height of the cavity inside the HPT-anvils is 0.6 mm, while the initial thickness of the samples is 0.8 mm. 
For comparison, the HPT-disk $N$0 which has only undergone compression and was not twisted in HPT-processing, is shown at $\varepsilon =$ 0. Clearly, there are two saturation stages for the increase of microhardness. 
When the strain is small, the hardness increases rapidly to about 525 HV. This might be due to the deformation by the compression in HPT to allow for the required shear strain. Subsequent shearing leads to a gradual increase to the first saturation of about 550 HV at a strain of around 70. 
Thereafter, the hardness shows a drastic increase to about 625 HV at a strain of about 100, followed by a slow increase and a tendency to a saturation at about 675 HV at a strain of $\varepsilon\sim$ 400. 
The first saturation stage is more clear when plotted over the logarithm of the applied strain as shown in Fig. \ref{fig:microhardness_log} in the appendix. 
\par

\subsubsection{Microstructural evolution}
\begin{figure}[!ht]
    \centering
    \includegraphics[width=\textwidth]{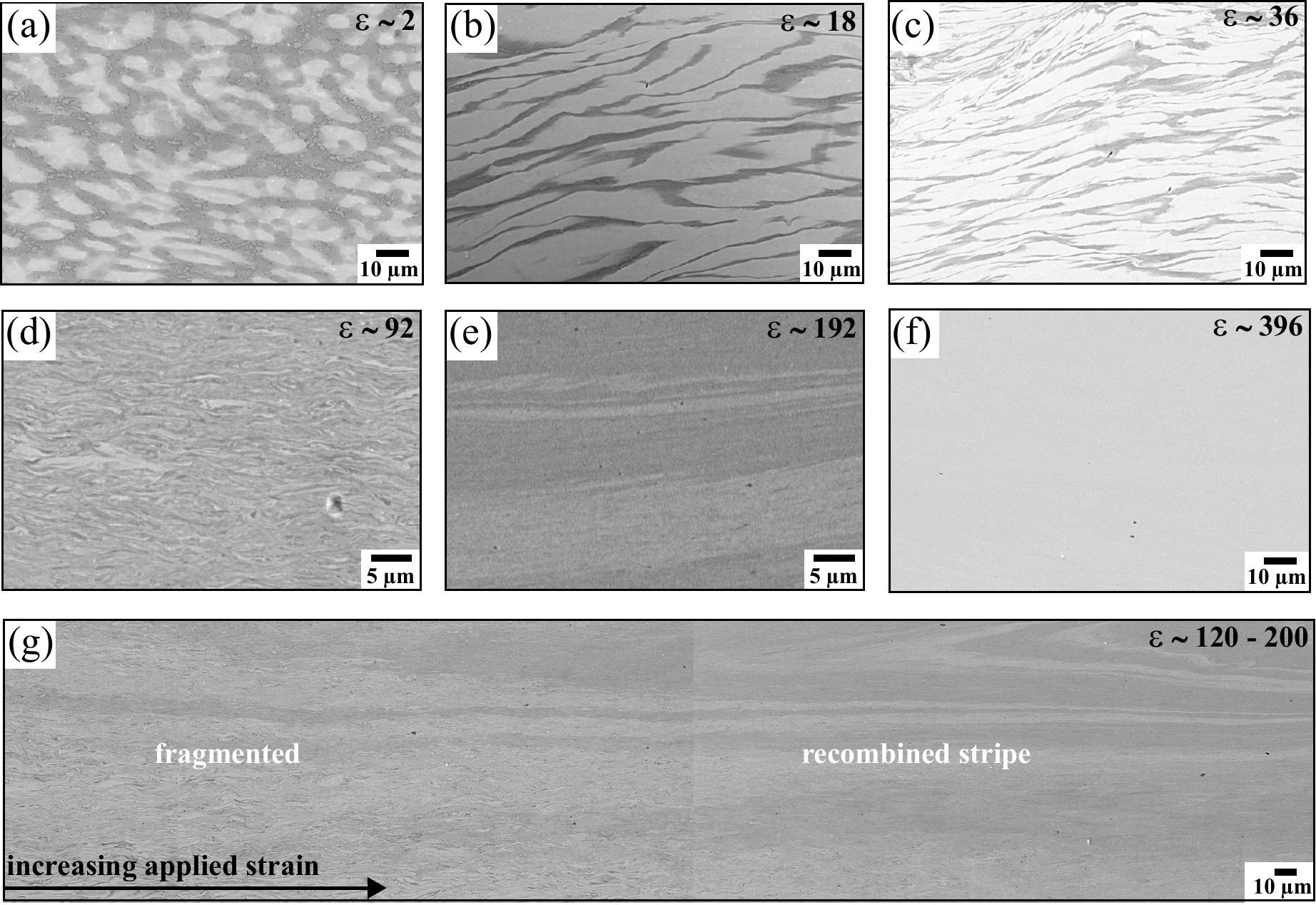}
    \caption{SEM micrographs show the microstructure at varied $\varepsilon$: (a) $\sim$2, (b) $\sim$18, (c) $\sim$36, (d) $\sim$92, (e) $\sim$192, (f) $\sim$396 and (g) 120 - 200. }
    \label{fig:SEM}
\end{figure}

The secondary electron (SE) images showing the microstructure at varied strains are presented in Fig. \ref{fig:SEM}. In our configuration, the SE detector in the SEM also receives signals from back scattered electrons, and thus the Z-contrast is shown as well. 
\par
With increasing shear strain, a slightly deformed dendritic microstructure (Fig. \ref{fig:SEM}a) evolves to a lamellar microstructure containing elongated dendritic and interdendritic regions (Fig. \ref{fig:SEM}b) at $\varepsilon\sim$ 18. Such a lamellar microstructure fragments and becomes thinner with further increasing strain. The thickness of the lamella reduced from $\sim$10 \textmu m at the strain of $\varepsilon\sim$ 18 to about 1 \textmu m at the strain of $\varepsilon\sim$ 92, leading to the formation of fragmented vortex microstructure, as is shown in Figs. \ref{fig:SEM}b-d. The vortex microstructure evolves to a continuous stripe-like microstructure when the strain is further increased to 192 (Fig. \ref{fig:SEM}e). A homogeneous microstructure is observed at a strain of $\varepsilon\sim$ 396 (Fig. \ref{fig:SEM}f). Fig. \ref{fig:SEM}g shows an overview of the evolution from the fragmented vortex to the stripe-like microstructure at $\varepsilon\sim$ 120 to 200. Note that the micrographs of Figs. \ref{fig:SEM}a-f are taken at $r$ = 1.5 mm for $N$0.25, $N$2, $N$4, $N$10, $N$20 and $N$40, respectively. Fig. \ref{fig:SEM}g is combined from 5 micrographs along the radius of specimen $N$20 at about $r=$ 1 - 1.5 mm. 
\par
The microhardness values corresponding to the characteristic microstructure of Figs. \ref{fig:SEM}a-f are listed in Tab. \ref{tab:microhardness}. It is reasonable to conclude that the microhardness increases by 50 HV, while the microstructure changes from dendritic to elongated lamella. The microhardness remains almost the same during the elongation process of the dendritic microstructure until it begins to fragment. The fragmentation leads to an increase in hardness, and the hardness of the stripe-like microstructure at $\varepsilon\sim$ 192 is about 3\% higher compared to that at $\varepsilon\sim$ 92. When the microstructure becomes chemically homogeneous, microhardness saturates. 
\par

\begin{table}[!ht]
	\centering
	\caption{Microhardness $\mathrm{HV}_{0.1}$ of the microstructures shown in Figs. \ref{fig:SEM}a-f.}
	\begin{tabular}{c c c c c c c}\hline
		$\varepsilon$ & $\sim$2 & $\sim$18 & $\sim$36 & $\sim$92 & $\sim$192 & $\sim$396 \\ \hline
		$\mathrm{HV}_{0.1}$ & {526.6$\pm$17.1} & {559.1$\pm$19.1} & {562.6$\pm$8.5} & {616.1$\pm$11.9} & {635.7$\pm$6.1} & {662.1$\pm$9.4}	 \\ \hline
	\end{tabular}
	\label{tab:microhardness}
\end{table}

%------------------------------XRD-----------------------------

\begin{figure}[!ht]
	\centering
	\includegraphics[width=\textwidth]{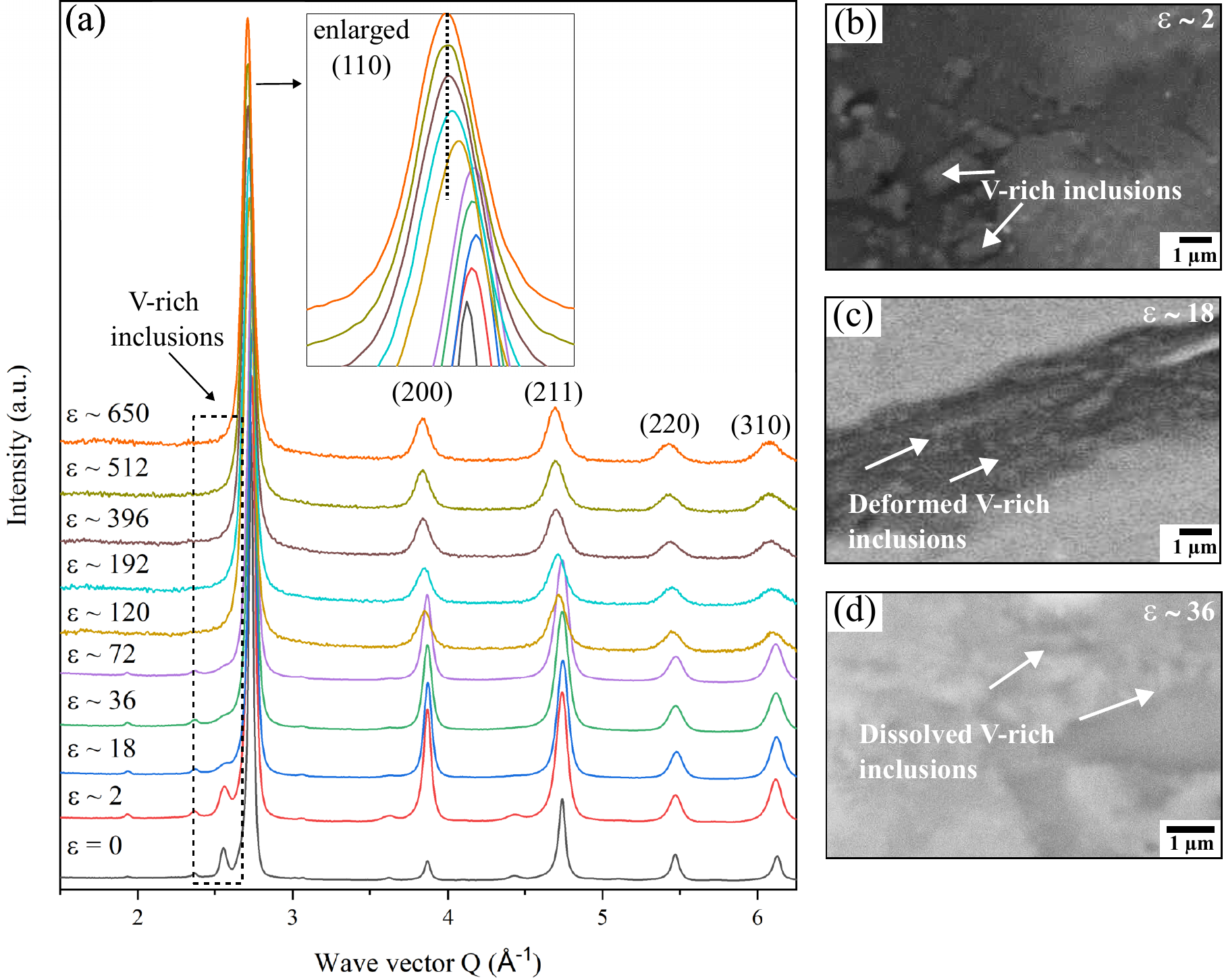}
	\caption{(a) X-ray diffractograms at different strains $\varepsilon$. (b-d) SEM micrographs show the deformation and fragmentation of the V-rich inclusions at $\varepsilon\sim$ 2, 18 and 36, respectively. }
	\label{fig:XRD}
\end{figure} 
In order to check the phase evolution during HPT-processing, the X-ray diffractograms measured at different applied strain $\varepsilon$ are depicted in Fig. \ref{fig:XRD}a. 
The $\varepsilon$ values in the figure are calculated with the position of the beam center taken as $r$ in Eq. \ref{eq:epsilon}. 
The diffractogram for $\varepsilon=$ 0 refers to the result obtained from specimen $N$0, which is merely compressed, but not twisted in the processing. 
\par 
It can be seen that the major BCC peaks are present throughout the HPT-processing, while the minor peaks representing V-rich inclusions gradually disappear. 
In addition, it can be found that the major BCC peaks broaden at the beginning of shearing when $\varepsilon$ increases from 0 to 2. The broadening is not obvious from $\varepsilon\sim$ 2 to $\varepsilon\sim$ 36. However, a severe broadening of the peak of the V-inlusions is observed in this strain range, which indicates the large deformation of the V-rich inclusions, as is proved by the micrographs taken at high magnification, see Figs. \ref{fig:XRD}b-d. 
The peak of the V-rich inclusions almost disappears after $\varepsilon$ increases to 72, indicating a dissolution of V-rich inclusions. 
A significant peak broadening of the major BCC peaks and the disappearance of V-rich inclusions can be observed when $\varepsilon$ increases from 72 to 120. Thereafter, a further pronounced broadening is not observed up to $\varepsilon\sim$ 650. 
Note that the XRD patterns from $\varepsilon\sim$ 120 onward were measured using CuK$\alpha$ radiation due to limited synchrotron X-ray beam time. Despite of that we found in a calibration experiment that the effect of instrumental broadening was negligible compared to that introduced by HPT-processing. 
The inset of Fig. \ref{fig:XRD}a shows the shift of the (110)-peak of the major phase. 
A slight peak shift to larger wave vectors Q can be found up to $\varepsilon\sim$ 18, and then to smaller Q up to $\varepsilon\sim$ 650. 
The peak shift suggests a decrease of the lattice parameter up to $\varepsilon\sim$ 18 and a subsequent increase up to $\varepsilon\sim$ 396. 
Interestingly, the (110)-peak shift is observed after the applied strain $\varepsilon$ reaches 120 and 192, at which the V-rich inclusions dissolve and only fragmented dendritic and interdendritic regions exist. 
\par

\begin{figure}
	\begin{center}
		\includegraphics[width=\textwidth]{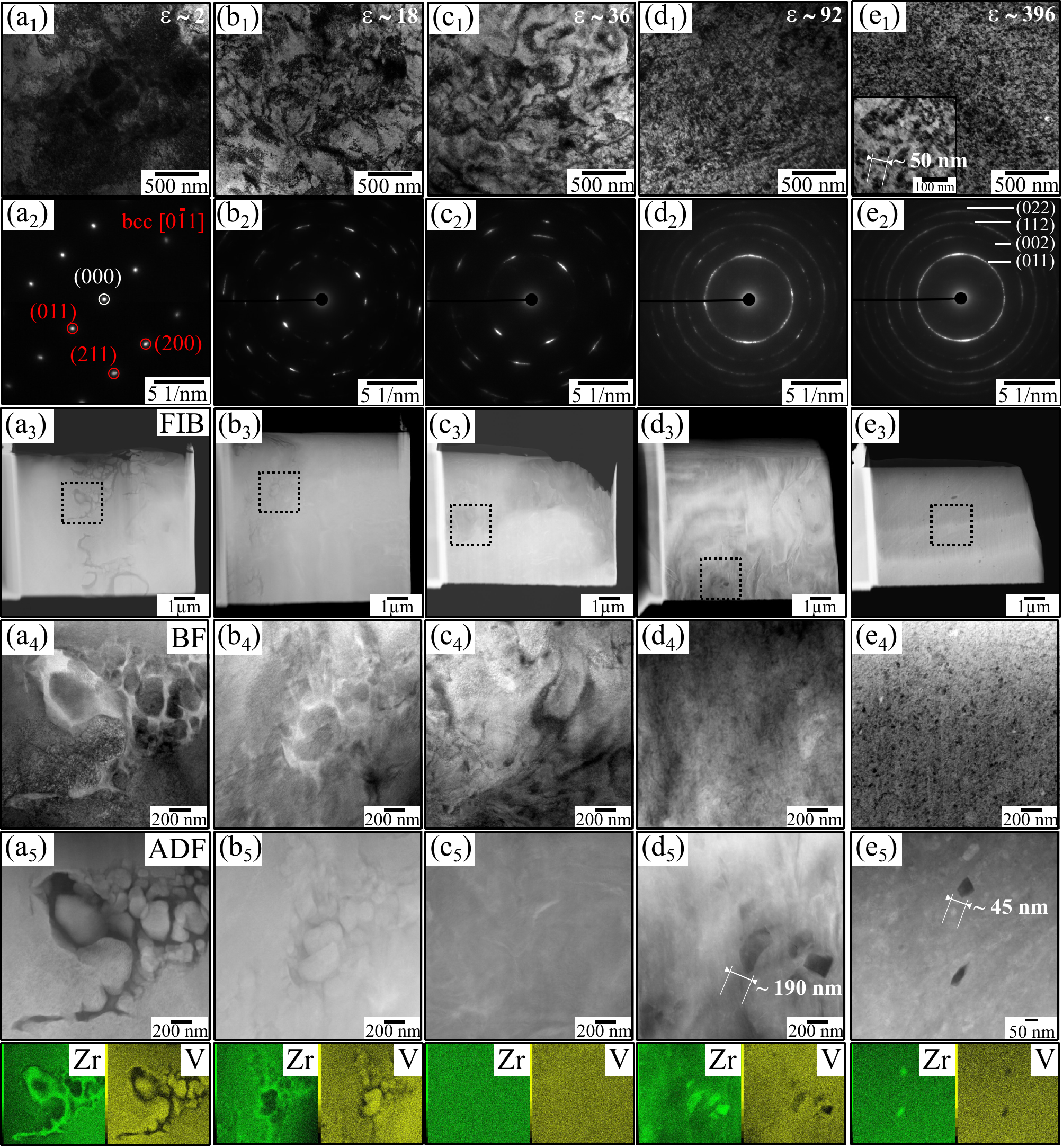}
		\caption{TEM investigations at $\varepsilon\sim$ (a) 2, (b) 18, (c) 36, (d) 92 and (e) 396. Bright field (BF) images ($_1$) and corresponding SAED patterns ($_2$). Low magnification ADF STEM showing the FIB lift-outs ($_3$), BF STEM  images ($_4$) and ADF images ($_5$) along with Zr EDS-maps and V EDS-maps at the bottom. The region of interest in the FIB lift-outs ($_3$) is indicated by a rectangle.}
		\label{fig:TEM}  
	\end{center}
\end{figure} 

Results of TEM investigations at $\varepsilon \sim$ 2, 18, 36, 92 and 396 are depicted in Fig. \ref{fig:TEM}. The FIB lift-outs were extracted at $r=1.5$ mm of the specimens $N$0.25, $N$2, $N$4, $N$10 and $N$40, respectively. 
\par
All bright field (BF) TEM micrographs shown in the Fig. \ref{fig:TEM} reveal very strong contrast related with the high density of defects like dislocations and grain boundaries. The pronounced dark contrast in Fig. \ref{fig:TEM}a$_1$ is additionally enhanced by the strong Bragg crystallographic orientation contrast. The corresponding selected area electron diffraction (SAED) (Fig. \ref{fig:TEM}a$_2$) shows a typical BCC spot pattern. The scanning transmission electron microscopy (STEM) BF image (Fig. \ref{fig:TEM}a$_4$) and the angular-dark-field (ADF) image (Fig. \ref{fig:TEM}a$_5$) clearly show the dendritic/interdendritic regions. After strain increases to $\varepsilon\sim$ 18, the diffraction spots become elongated and diffused (Fig. \ref{fig:TEM}b$_2$) which is driven by the accumulated distortion of the crystal lattice. This effect is even more pronounced for $\varepsilon\sim$ 36 (Fig. \ref{fig:TEM}c$_1$). The STEM BF and ADF micrographs for $\varepsilon\sim$ 18 (Figs. \ref{fig:TEM}b$_4$ and b$_5$) show the presence of V-rich inclusions. Only after applying higher strains the V-rich inclusions dissolve (Fig. \ref{fig:TEM}c$_4$, c$_5$). The BF TEM micrograph for $\varepsilon\sim$ 92 (Fig. \ref{fig:TEM}d$_1$) reveal the presence of an ultra-fine grained (UFG) microstructure. The corresponding SAED pattern (Fig. \ref{fig:TEM}d$_2$) exhibits rings (identical selected area aperture is used in acquisition of all diffraction patterns). At this applied strain, the coarse dendritic and interdendritic regions are not distinguishable. Few remaining Zr-rich particles with a size of about 190 nm still can be found (Fig. \ref{fig:TEM}d$_5$). A further increase of the HPT strain to $\varepsilon\sim$ 396 results in further structure refinement, as shown in Fig. \ref{fig:TEM}e$_1$ and e$_4$. However, the identical SAED patterns at $\varepsilon\sim$ 92 and $\sim$396 (\ref{fig:TEM}d$_2$ and e$_2$) indicate such a refinement is not intensive. Please note that in Fig. \ref{fig:TEM}d$_1$ and e$_1$, the scale bar is the same, as is the case for Fig. \ref{fig:TEM}d$_4$ and e$_4$. An enlarged view of Fig. \ref{fig:TEM}e$_1$ indicates that the size of the grains is about 50 nm. Similar to $\varepsilon\sim$ 92, few small Zr-rich particles can be found undissolved, yet with a smaller size of about 45 nm. 
\section{Discussion}
\label{dissc}
\begin{figure}[!ht]
	\centering
	\includegraphics[width=\textwidth]{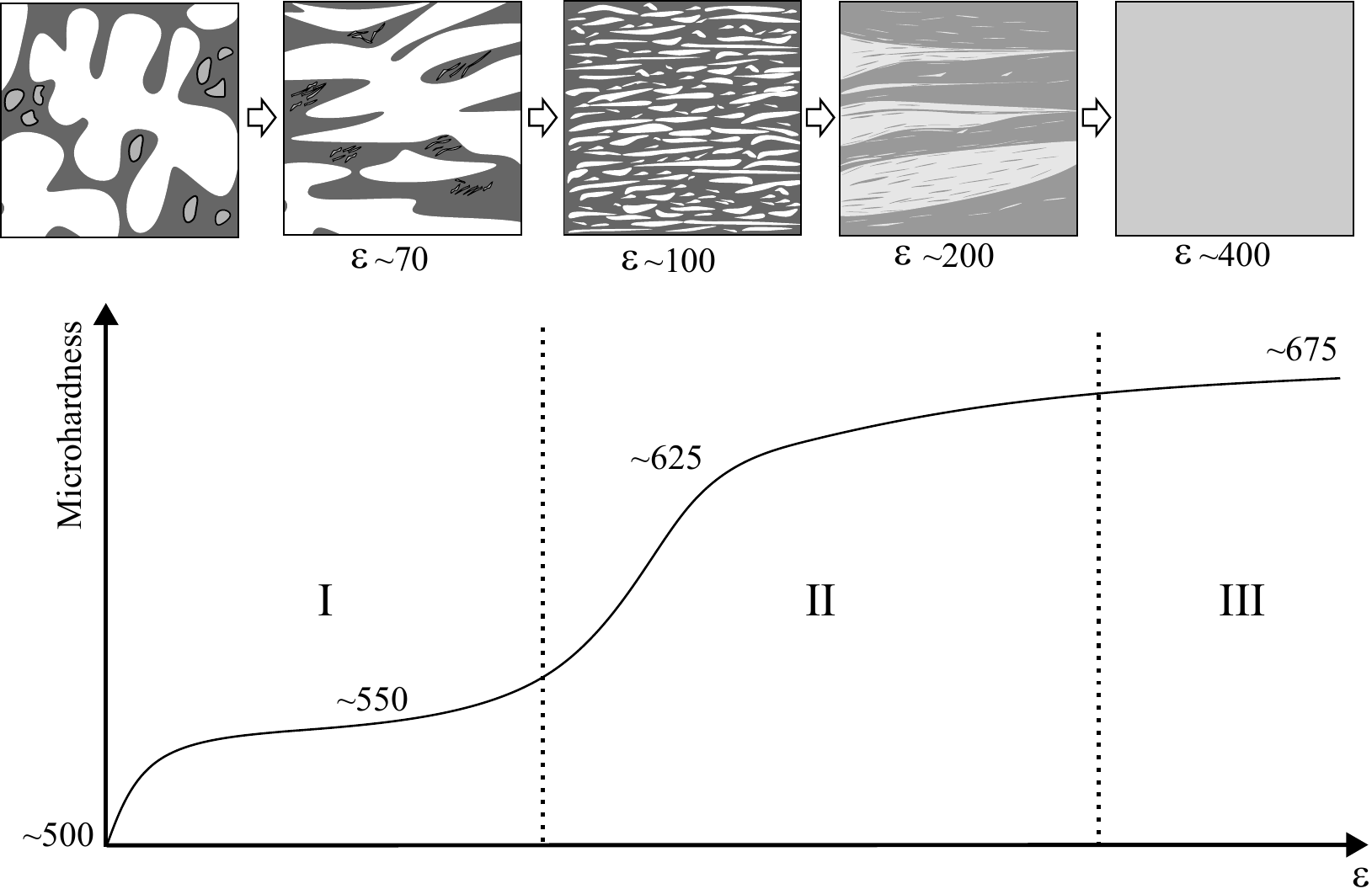}
	\caption{Schematic illustration of microstructure and its corresponding microhardness evolution in the three deformation stages. }
	\label{fig:Sketch}
\end{figure} 

Fig. \ref{fig:Sketch} provides a summary of the evolution of the previously described microstructure and the corresponding microhardness. In the paragraphs that follow, we will discuss the deformation mechanisms and explain the reasons for the enhanced hardness. 

\subsection{Deformation mechanism}
\subsubsection{Stage I ($\varepsilon\sim$ 0 to $\sim$ 70)} 
\par
\begin{figure}[!ht]
	\centering
	\includegraphics[width=\textwidth]{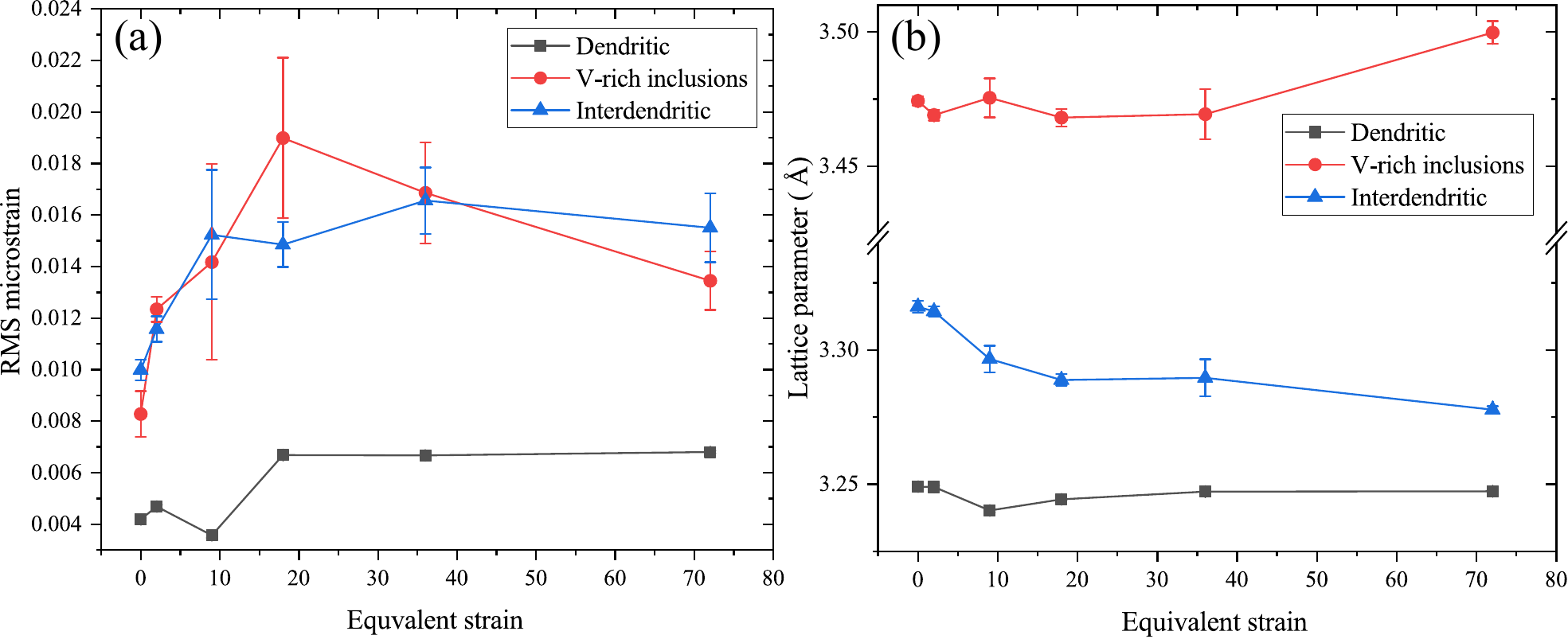}
	\caption{Evolution of (a) root mean square (RMS) microstrain and (b) Lattice parameter obtained from Rietveld refinement in the first deformation stage.}
	\label{fig:refinement}
\end{figure} 
In XRD peak broadening, the effect of strain can be mathematically separated from that of crystallite/diffracting domain size, assuming that the broadening due to strain is Gaussian and the effect of size is Lorentzian \cite{Young1993}. 
By Rietveld refinement of the synchrotron XRD results up to $\varepsilon\sim$ 72, we obtain the evolution of the root mean square (RMS) microstrain and lattice parameter during the first deformation stage, as shown in Figs. \ref{fig:refinement}a and b, respectively. 
It is obvious that the RMS strains in the interdendritic region (blue) and in the V-rich inclusions (red) are much larger than that in the dendritic region (black), indicating a strain localization in the interdendritic region. Such a strain localization might be due to the large deformation of V-rich inclusions. Since in our previous study \cite{Duan2022} no pronounced enrichment of vanadium is found in the dendritic region, it can be assumed that the V-rich inclusions are mainly dissolved in the interdendritic region. 
This assumption could be confirmed by the lattice parameter of the interdendritic region in Fig. \ref{fig:refinement}b. Since V has a smaller atomic radius than the other elements in the RHEA, the lattice parameter of the interdendritic region should become smaller when V is dissolved in its lattice. It is also noted that the lattice parameter of the V-rich inclusions shows an increase between $\varepsilon\sim$ 36 and $\varepsilon\sim$ 72. Since multiple phases are present in the RHEA and since HPT is a mechanical process that introduces high shear strains, the change in lattice constants (increase in the V-rich inclusions and decrease in the interdendritic region) might suggest the presence of intergranular stresses between different phases. Nevertheless, the relationship between the formation of residual stress and HPT in the RHEA has not been studied in depth, so no conclusive inference can be made about such phenomenon.
\par
The deformation mechanism of multi-phase materials in HPT-processing is usually much more complex than that of single-phase materials. In the study of binary immiscible composites processed by HPT, the strain localization is a consequence of the hardness difference between the phases \cite{Kormout2017}. This is due to the fact that the soft phase is more prone to deformation compared to the hard phase in HPT-processing. In our previous study \cite{Duan2022}, a strain softening behavior was observed by probing the average ultra-microhardness of the interdendritic region. As shown in Figs. \ref{fig:XRD}b-d, the deformed V-inclusions are distributed in the interdendritic region. It is reasonable to conclude that the V-inclusions are softer and lead to a strain localization in the interdendritic region. This strain localization and the dissolution of the V-rich inclusion could explain the moderate hardness increase during deformation stage I. A similar phenomenon was observed in the study of the supersaturated solid solution of the Cu-Cr system processed by HPT \cite{Bachmaier2014}. Bachmaier et al. attributed this to the refinement and saturation of hardness within the respective phases, analogous to the behavior of pure metals \cite{Bachmaier2014}. This is consistent with the observation in our work, as we did not find significant grain refinement in the strain range from $\varepsilon\sim$ 18 to 36 (Figs. \ref{fig:TEM}b and c). 

\subsubsection{Stage II ($\varepsilon\sim$ 72 to $\sim$ 300)} 
\begin{figure}[!ht]
	\centering
	\includegraphics[width=\textwidth]{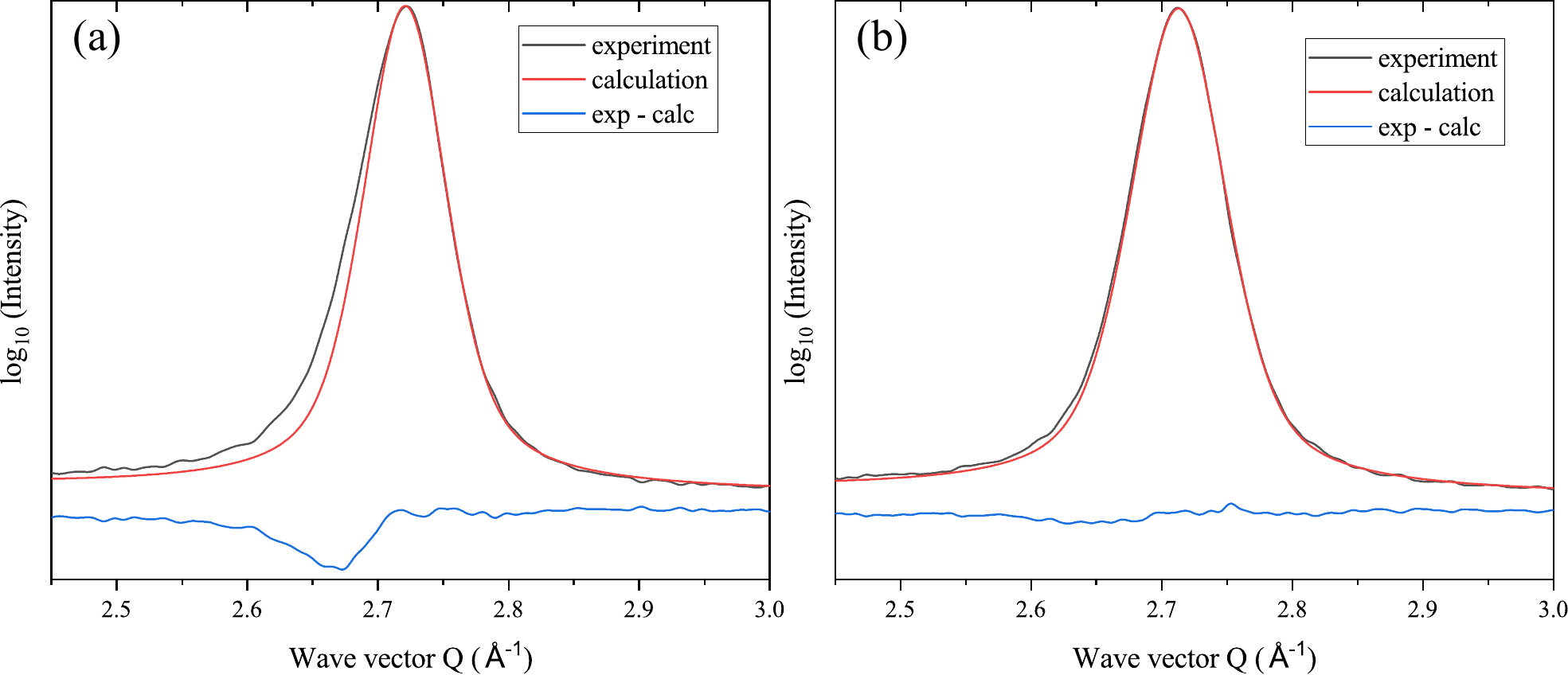}
	\caption{Enlarged view of the (110)-peak in XRD at (a) $\varepsilon\sim$ 92 and (b) $\varepsilon\sim$ 396, assuming there exists only one BCC phase. }
	\label{fig:Asym2}
\end{figure} 
After the dissolution of the V-rich inclusions, the main deformation mechanism is the co-deformation of the dendritic and interdendritic regions. In deformation stage II, fragmentation as well as the recovery of the phases compete with each other to finally reach an equilibrium \cite{Pippan2010}. As shown in Fig. \ref{fig:SEM}d, the elongated dendritic region is fragmented at $\varepsilon\sim$ 92. 
The (110)-peak is still asymmetric (Fig. \ref{fig:Asym2}a), which disappears after the microstructure is fully homogenized, e.g. at $\varepsilon\sim$ 396 (Fig. \ref{fig:Asym2}b). 
Again, this demonstrates that the interdendritic phase exists. 
The peaks of the two phases are so close to each other that the errors in the results obtained from Rietveld refinement are too large to be reliable. 
\par
\begin{figure}[!ht]
	\centering
	\includegraphics[width=\textwidth]{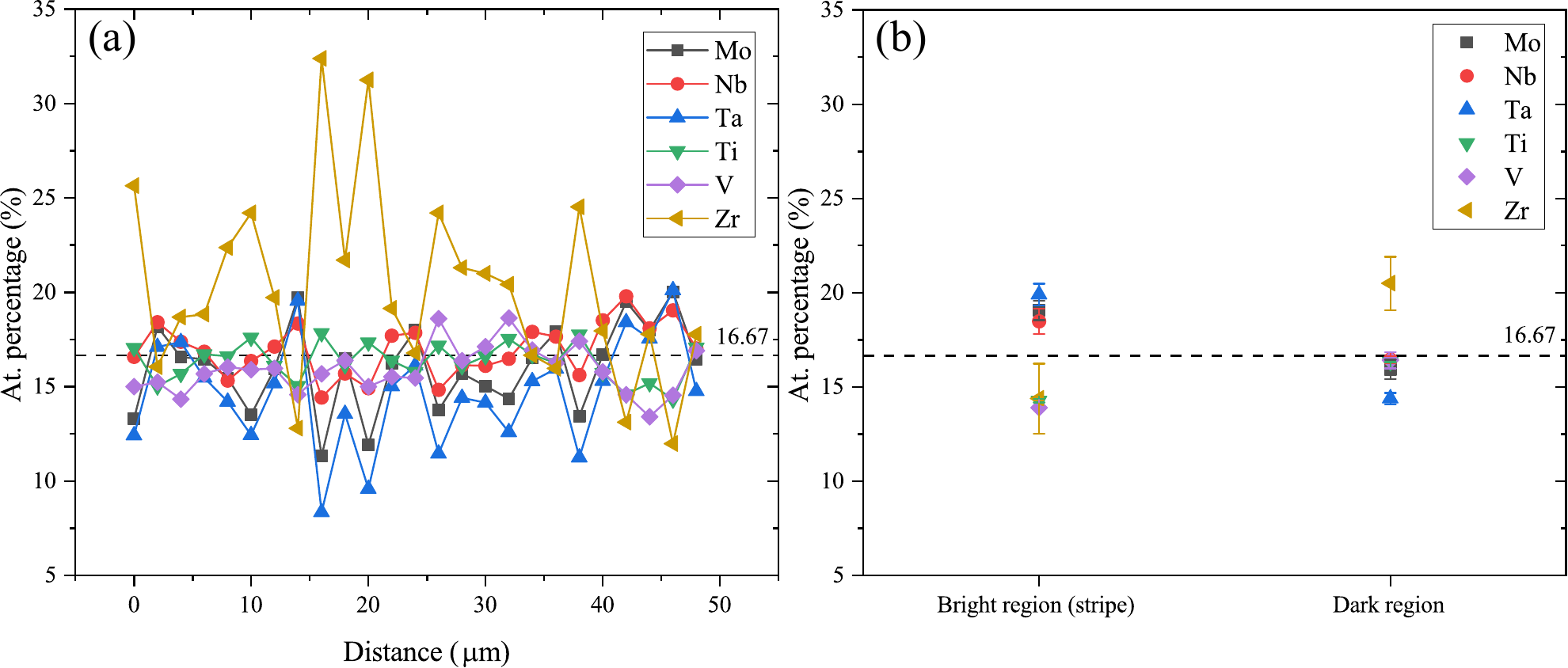}
	\caption{(a) An EDS-line scan with a measurement distance of 2 \textmu m at $\varepsilon\sim$ 92. (b) EDS-point analysis of the stripe and the matrix microstructure at $\varepsilon\sim$ 192. }
	\label{fig:EDS_dis}
\end{figure} 
Although the fragmented phases are already too small at this deformation stage to be measured accurately by EDS in the SEM, we can still roughly estimate the chemical composition by line-scans as shown in Fig. \ref{fig:EDS_dis}a. Apparently, there is a significant segregation of Zr (brown) and Ta (blue) compared to the theoretical value of 16.67\% for the senary HEA. The segregation of Zr is more pronounced than that of Ta. It can be seen that the Zr-concentration exceeds 30 at.\% in some Zr-rich areas. Note that the observed segregation should be less than the actual one due to the large interaction volume of SEM-EDS. Up to $\varepsilon\sim$ 92, it can be assumed that the shear-induced mechanical mixing between the dendritic and the interdendritic regions might be negligible. At $\varepsilon\sim$ 192, the width of the recombined stripe is larger than 5 \textmu m. Hence, the quantitative EDS-analysis on the microstructure is more accurate (Fig. \ref{fig:EDS_dis}b). We find that the elemental segregation is already very small at this applied strain, indicating that the mixing of elements is much faster during shearing from $\varepsilon\sim$ 100-200 compared to $\varepsilon\sim$ 0-100. 
\par
Various mechanisms have been proposed to explain the deformation-induced mechanical mixing in immiscible system. Bellon and Averback suggested a kinetic roughening of interfaces leading to the forced formation of a solid solution \cite{Bellon1995}. In addition, Yavari et al. proposed the Gibbs-Thomson effect of the fragmented phases after reaching a critical size (below 1 nm), they could then be dissolved into the other phase due to capillary pressure \cite{Yavari1992}. 
More recently, a dislocations shuffling mechanism was proposed, where dislocations glide across the heterophase interfaces and carry atoms from one phase to the other \cite{Raabe2009,Raabe2010}. It was also suggested that one phase would eventually dissolve into the other by a process comparable to erosion or abrasion \cite{Bachmaier2016}. 
For the mechanical mixing of the dendritic and interdendritic regions of the HEA in the present study, it is more resonable to conjecture that the erosion and abrasion process proposed in \cite{Bachmaier2016} could explain this phenomenon. After all, we do observe fragmented Zr-rich particles in the TEM analysis. 
\subsection{On the "single-phase" dendritic RHEA and its homogenization}
The determination of the single-phase composition in an HEA is usually based on the assumption that peaks of only one phase are found in the XRD pattern. Hence, although some HEAs showed deviations from chemical homogeneity, such as segregation or dendrites, they were still considered to be single-phase \cite{Gao2016,Senkov2010}. A recent study has shown that the "single-phase" HEAs containing elemental segregation are actually composed of two phases with the same crystal structure and very close lattice constants \cite{Zhang2017}. The results obtained in this work clearly support the latter opinion. HEAs are usually composed of elements with similar physical properties such as atomic radii and electronic concentration. When multiple elements are grouped together in an HEA, the interactions between them are complex, such as a positive enthalpy of mixing that leads to a miscibility gap. It is conceivable that they have similar lattice constants after phase decomposition such that they cannot be distinguished from each other in XRD. 
\par
The common dendritic microstructure in RHEAs was reported to have a high thermal stability \cite{Wang2021,Li2022}. It is assumed that the solution annealing temperature of RHEAs is extremely high and the homogenization heat treatment could be impractical and uneconomical. 
In particular, RHEAs containing both Ta and Zr are susceptible to segregation and are forming dendrites during solidification \cite{Gao2016,Senkov2021,Senkov2021a,Zyka2019,Whitfield2020}. Zyka et al. attributed this to the entropy effect from their studies of ternary or quaternary RHEAs \cite{Zyka2019}. In the present study, dendrites are also present with six equimolar elements. The interdendritic region is likely to be rich in Zr even if there exists an element with lower melting point such as Ti \cite{Gao2016,Senkov2021}. This is likely due to the fact that the liquidus temperature of Ta$_{0.5}$Zr$_{0.5}$ is 800 $^{\circ}$C lower than that of Ta$_{0.5}$Ti$_{0.5}$, and the liquid phase would be stabilized by Zr during solidification \cite{Gao2016}. 
Whitfield et al. found the dendritic microstructure in the ternary Ta-Ti-Zr system was still observable after annealing at 1200 $^{\circ}$C for 100 hours followed by water quenching, and the dendrites could be only be eliminated by an additional heat treatment at 1000 $^{\circ}$C for 1000 hours again followed by water quenching \cite{Whitfield2020}. 
For the RHEA in the present study, an annealing of the as-solidified sample at 1400 $^{\circ}$C for 12 hours followed by cooling in the furnace did not result in chemical homogeneity. A subsequent annealing at 1000 $^{\circ}$C for 12 hours followed by quenching did not homogenize the RHEA either (see Fig. \ref{fig:HeatTreated} in the appendix). 
However, the results of our study show that the homogenization of RHEA can be achieved by mechanical processing. Moreover, while achieving chemical homogeneity, the grains are also refined and enhanced mechanical properties are obtained. 
\par
\subsection{The enhanced microhardness}
The yield strength $\sigma_{\mathrm{y}}$ of a metallic material can be expressed as the sum of the intrinsic lattice friction stress ($\sigma_{\mathrm{fr}}$) and contributions of different strengthening mechanism \cite{Wu2014a}: 
\begin{equation}
	\sigma_{\mathrm{y}} = \sigma_{\mathrm{fr}} + \Delta\sigma_{\mathrm{\rho}\mathrm{i}} + \Delta\sigma_{\mathrm{ss}} + \Delta\sigma_{\mathrm{ppt}} + \Delta\sigma_{\mathrm{gb}}, 
	\label{eq:yieldstrength}
\end{equation}
\par
where the terms on the right are the contributions of the initial dislocation density $\Delta\sigma_{\mathrm{\rho}\mathrm{i}}$, solid solution hardening $\Delta\sigma_{\mathrm{ss}}$, precipitation hardening $\Delta\sigma_{\mathrm{ppt}}$ and grain boundaries $\Delta\sigma_{\mathrm{gb}}$. Obviously, in the present study, there is no contribution of precipitation hardening $\Delta\sigma_{\mathrm{ppt}}$. The first increase in hardness, up to $\varepsilon\sim$ 10, is due to the increase of the dislocation density $\Delta\sigma_{\mathrm{\rho}\mathrm{i}}$ (Fig. 5a$_1$) and the grain refinement within the dendritic and the interdendritic regions $\Delta\sigma_{\mathrm{gb}}$. The Peierls stress in BCC metals is much higher compared to that in FCC metals. A recent study has shown that the compositional randomness in BCC high entropy alloys could additionally increase the Peiers stress, resulting in an extremely high intrinsic strength $\sigma_{\mathrm{fr}}$ \cite{zhang2019effect}. This might explain why the increase of hardness in stage I is not significant compared to the initial hardness. The dramatic increase in hardness at about $\varepsilon\sim$ 92 (Fig. 6) should be attributed to the grain refinement by an intense fragmentation. As suggested by Bachmaier et al. \cite{Bachmaier2014}, the decrease of the mean inter-particle distance (referred to as "phase spacing" in their study) in the Cu-Cr system could further refine the grains that are already saturated. Since we did not find intense grain refinement from $\varepsilon\sim$ 92 to $\varepsilon\sim$ 396 from the TEM results, the subsequent increase in hardness can only be explained by the additional solid solution hardening $\Delta\sigma_{\mathrm{ss}}$ induced by the dissolution of the large atom Zr. Interestingly, the increase in hardness due to grain refinement ($\sim$ 75 HV) in stage II is not much different from that due to solid solution hardening ($\sim$ 50 HV). 
\par
Severe lattice distortion is considered to be one of the core effects in HEAs \cite{Miracle2017}. Among the various elements that form HEA, zirconium, which possesses large atomic radii, has been suggested to particularly cause local chemical fluctuations and significant lattice distortion \cite{Tong2020a}. Although the results in the present investigation also support this opinion, the fourteen RHEAs studied in \cite{Tong2020a} were all dendritic. There still exists some doubt if better mechanical properties can be achieved when the microstructure is homogenized. 

\section{Conclusions}
\label{concl}
The dendritic senary RHEA MoNbTaTiVZr was processed by HPT up to 40 revolutions. The microstructural evolution was comprehensively studied by SEM, XRD and TEM. The correlation between enhanced hardness and microstructure, as well as the deformation and mixing mechanisms were discussed. In summary, the following conclusions can be drawn:
\begin{itemize}
    \item The dendritic RHEA is found to achieve chemical homogenization with simultaneously enhanced hardness by HPT-processing. 
	\item The deformation and mixing mechanism can be analogous to the mechanical mixing in binary immiscible systems. 
	\item The hardness enhancement at low strains (up to $\varepsilon\sim$ 10) is mainly due to the increase in dislocation density and grain refinement within the respective phase. This increase is unusually insignificant (50 HV) compared to the initial hardness (500 HV). The reason might be the high Peierls stress in BCC HEA. 
	\item At intermediate strains ($\varepsilon\sim$ 100), the dramatic hardness increase (75 HV) is attributed to further grain refinement caused by fragmentation. An UFG microstructure is formed. 
	\item An UFG microstructure is present at $\varepsilon\sim$ 400, where the Zr-depleted dendritic and the Zr-rich interdendritic regions are mixed homogeneously. The hardness increase (50 HV) could be explained by the solid solution strengthening resulting from the dissolution of the large Zr atom.

\end{itemize}
\section*{Acknowledgement}
The authors express their gratitude to DESY (Hamburg, Germany) for providing support and granting access to HEXRD facilities. Part of the experiments was conducted at PETRA III. The authors would also like to acknowledge the Materials Science Lab, which is operated by FRM II and Helmholtz-Zentrum Hereon at the Heinz Maier-Leibnitz Zentrum (MLZ), Garching, Germany, for preparing the samples used in this study. CD is grateful to Prof. Jan Torgersen for the financial support and insightful discussions during the writing of this work. CD thanks Mr. Patrick Hegele and Dr. Zhonghua Wang for their help with the XRD and SEM experiments. 
\bibliographystyle{elsarticle-num} 
\bibliography{MyBib}
\newpage
\appendix
\section{Influence of V on the microstructure of MoNbTaTiV$_{x}$Zr}
\label{sec:supplement1}
\begin{figure}[!ht]
	\centering
	\includegraphics[width=\textwidth]{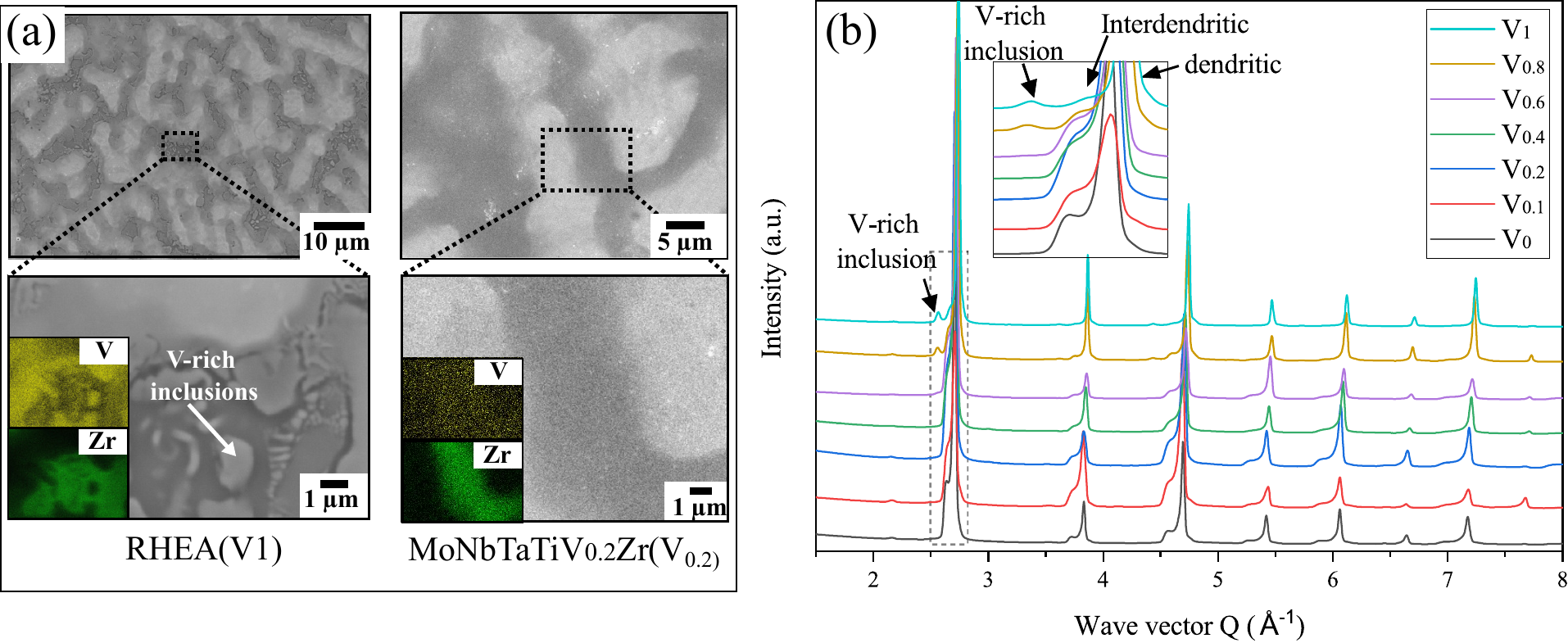}
	\caption{(a) Microstructure with V and Zr EDS-mapping results of MoNbTaTiV$_{1}$Zr (RHEA) and MoNbTaTiV$_{0.2}$Zr (V$_{0.2}$). (b) Synchrotron X-ray diffractograms of MoNbTaTiV$_{x}$Zr (V$_{x}$).}
	\label{fig:V_influence}
\end{figure} 
\newpage
\section{Asymmetry of the (110)-peak of the major phase as obtained by Rietveld refinement}
\begin{figure}[!ht]
	\centering
	\includegraphics[width=0.75\textwidth]{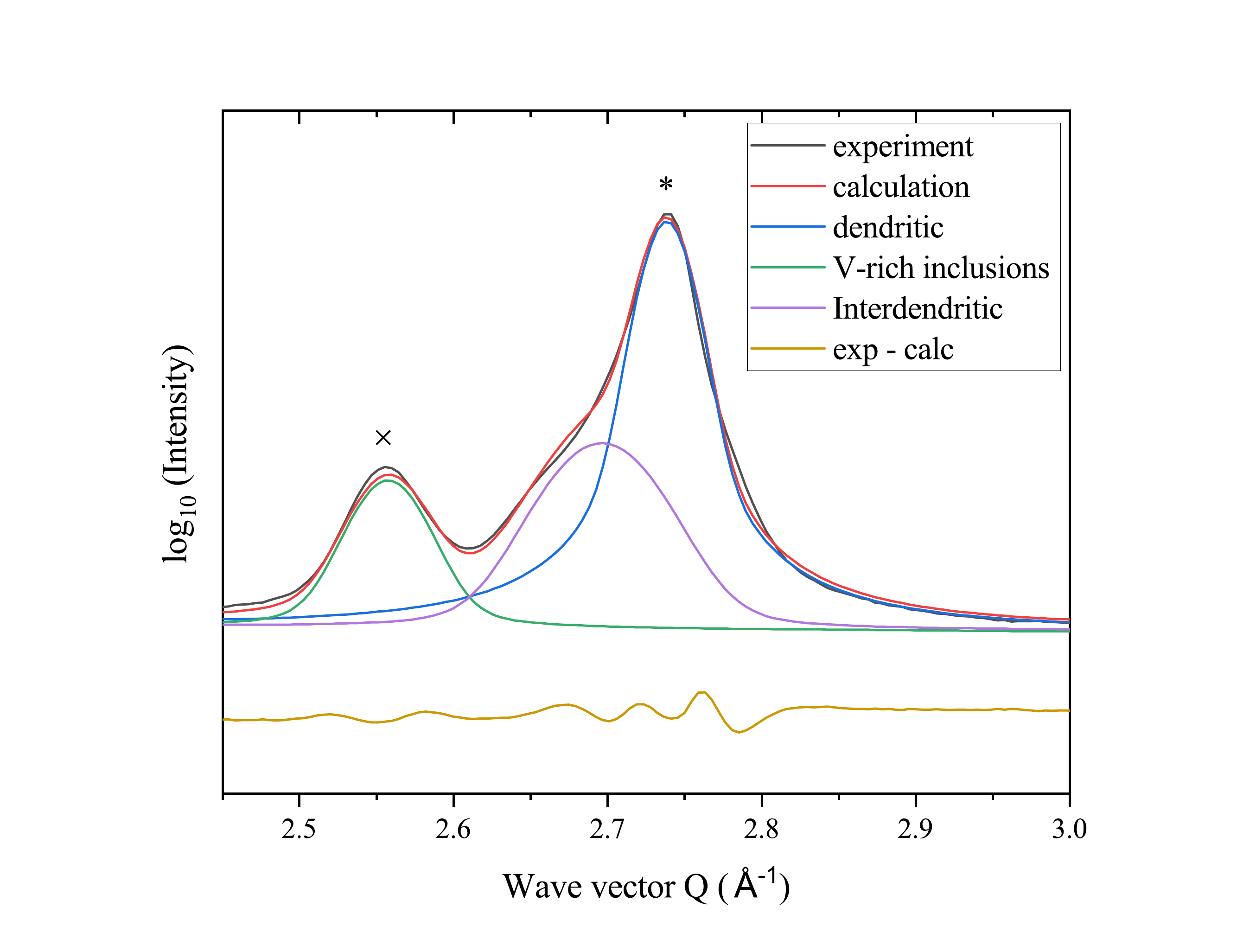}
	\caption{The enlarged view of the (110)-peak of RHEA in the Rietveld refinement.}
	\label{fig:AsymRR}
\end{figure} 
\newpage
\section{Plot of microhardness against the logarithm of the applied strain}
\par
\begin{figure}[!ht]
	\centering
	\includegraphics[width=0.75\textwidth]{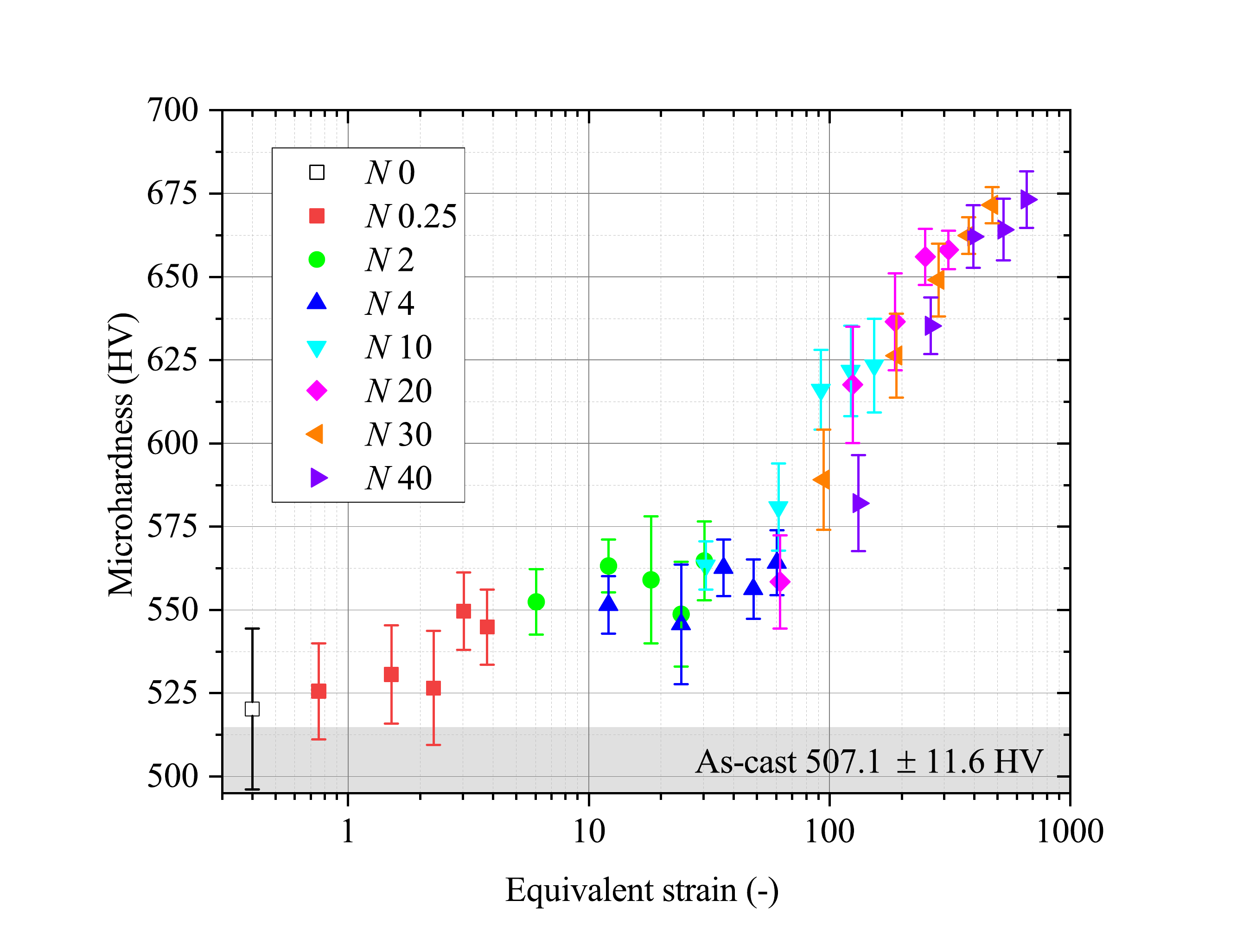}
	\caption{Microhardness evolution against logarithm of the applied strain during HPT-processing.}
	\label{fig:microhardness_log}
\end{figure} 
\section{Light optical microscopy image of the heat-treated RHEA}
\par
\begin{figure}[!ht]
	\centering
	\includegraphics[width=0.75\textwidth]{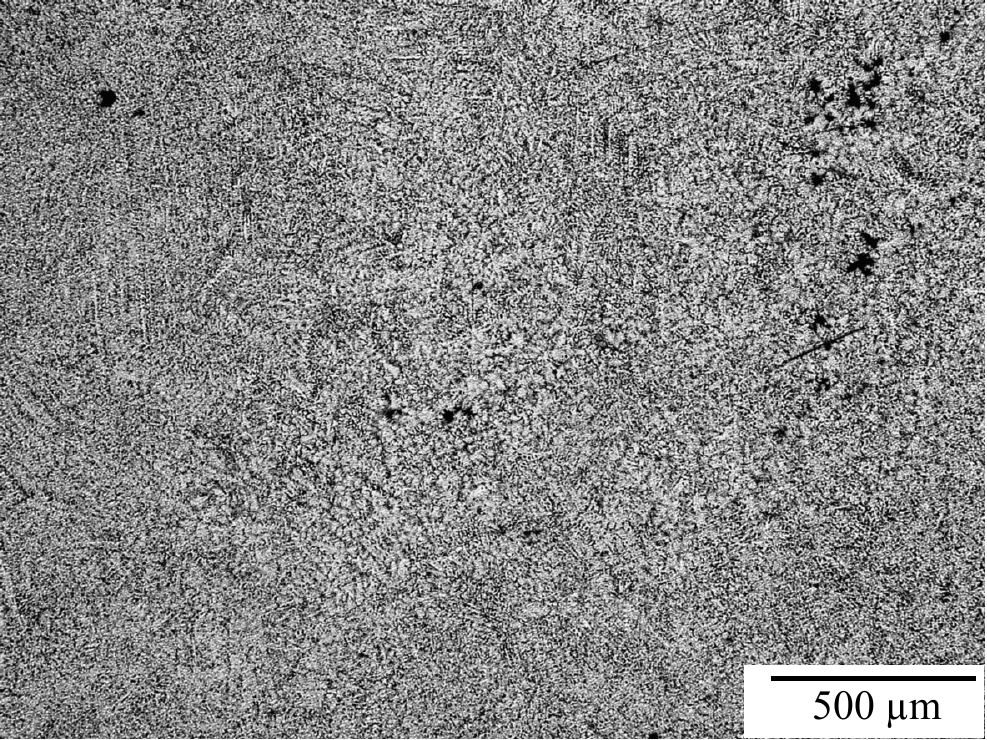}
	\caption{Light optical micrograph of the heat-treated RHEA. }
	\label{fig:HeatTreated}
\end{figure}

\end{document}